\documentclass[conference]{IEEEtran}
\IEEEoverridecommandlockouts

\usepackage{listings}
\lstdefinelanguage{SQL}{
  morekeywords={CREATE,TABLE,PRIMARY,KEY,CHECK,NOT,NULL,BOOLEAN,SMALLINT,BIGINT,VARBIT},
  sensitive=false,
  morecomment=[l]{--},    % line comment is --
  morestring=[b]',        % strings are '...'
}

\usepackage{tikz}
\usetikzlibrary{
  arrows.meta, 
  positioning, 
  shapes.geometric, 
  shapes.symbols, 
  fit, 
  calc
}
 
\usepackage{graphicx}
\usepackage{booktabs}

 \usepackage{cite}
\usepackage{amsmath,amsfonts}

\usepackage{algorithmic}
\usepackage{textcomp}
\usepackage{graphicx}

\usepackage{booktabs}
\usepackage{hyperref}
\usepackage{enumitem}

\usepackage{adjustbox}

\usepackage{multirow}
\usepackage{placeins}
\usepackage{outlines}
\usepackage[linesnumbered]{algorithm2e}
\usepackage{amsthm}
\usepackage{booktabs}
\usepackage{adjustbox}
\usepackage{blindtext}

\usepackage{ifsym}

\usepackage{braket}
\usepackage{mathtools}

\usepackage[utf8]{inputenc}
\usepackage[english]{babel}

\theoremstyle{definition}

\usepackage{tabularx}
\usepackage{hyperref}

\newcounter{RihanNOC}
\stepcounter{RihanNOC}

\newcounter{GiancarloNOC}
\stepcounter{GiancarloNOC}

 \newcounter{FlorisNOC}
\stepcounter{FlorisNOC}

\newcounter{td}

\newcounter{all}

\newcommand{\para}[1]{\noindent\textbf{#1.}}

\usepackage{tikz}
\usetikzlibrary{quantikz2}
 
\usepackage[framemethod=TikZ,
innerleftmargin=0.1\baselineskip]{mdframed}

\newcommand{\bemph}[1]{\textbf{\textit{#1}}}
\newmdenv[linecolor=white,backgroundcolor=mygray]{myframe}
 \definecolor{mygray}{gray}{0.95}
\newcommand*\colourcheck[1]{%
  \expandafter\newcommand\csname #1check\endcsname{\textcolor{#1}{\ding{52}}}%
}
\newcommand*\colourcross[1]{%
  \expandafter\newcommand\csname #1cross\endcsname{\textcolor{#1}{\ding{54}}}%
}

\definecolor{darkgreen}{rgb}{0.0, 0.2, 0.13}

\usepackage[hang,flushmargin]{footmisc} 

\def\BibTeX{{\rm B\kern-.05em{\sc i\kern-.025em b}\kern-.08em
    T\kern-.1667em\lower.7ex\hbox{E}\kern-.125emX}}

\begin{document}

% \title{Quantum Databases: the Ultimate Secure Databases}
% \title{Quantum Databases: the Ultimate Private Databases}
\title{Private Quantum Database}
 % NISQ-Ready Quantum Databases: Practical Design for User & Data Privacy

%%
%% The "author" command and its associated commands are used to define the authors and their affiliations.

% \author{Giancarlo Gatti}
% % \orcid{0000-0002-3720-6585}
% \affiliation{%
% \institution{   
% Mondragon University}
%   % \   \institution{Delft University of Technology}
%    \country{Spain}
% }
% \email{ggatti@mondragon.edu}
 
% \author{Rihan Hai}
% % \orcid{0000-0002-3720-6585}
% \affiliation{%
% \institution{   
% Delft University of Technology}
%   % \   \institution{Delft University of Technology}
%    \country{Netherlands}
% }
% \email{r.hai@tudelft.nl}
\author{
\IEEEauthorblockN{Giancarlo Gatti}
\IEEEauthorblockA{\textit{Mondragon University} \\
% Arrasate-Mondragón, Spain \\
ggatti@mondragon.edu}
\and
\IEEEauthorblockN{ Floris Geerts}
\IEEEauthorblockA{\textit{University of Antwerp} \\
% Antwerp, Belgium \\
floris.geerts@uantwerp.be}
\and
\IEEEauthorblockN{Rihan Hai}
\IEEEauthorblockA{\textit{Delft University of Technology} \\
% Delft, Netherlands \\
r.hai@tudelft.nl}
}

\maketitle

\begin{abstract}
We ask what a \emph{quantum-native} database can do for data management \emph{today}, and argue that two-sided privacy is a compelling near-term capability. We target single-server symmetric private information retrieval (SPIR), which requires both \emph{user privacy} (hiding which record is queried) and \emph{data privacy} (preventing retrieval beyond the requested record). Our design encodes relational tuples using quantum random access codes (QRACs), producing quantum states that embody the table’s contents. The server sends a bounded number of these states; the client makes one destructive measurement tied to the desired key, reconstructing only that tuple while making unqueried rows physically inaccessible due to measurement disturbance and basis incompatibility. To keep this deployable on current hardware, we propose a hybrid quantum–classical architecture: classical descriptions of the states live in a standard DBMS, and fresh instances are generated at query time. The result is single-server, SPIR-style privacy without trusted hardware or heavyweight cryptography, offering a concrete, application-driven path to introducing quantum components into future database systems.

\end{abstract}

\section{Introduction}
Database systems have been engineered for decades around classical data on classical machines. As quantum computing moves from laboratory curiosity to a commercially funded technology stack, our field faces a foundational question:
\begin{center}
    \emph{What can a quantum database do?}
\end{center}
Answering this requires application‑driven analysis and realistic system design, while engaging honestly with the limitations and potential of today’s quantum hardware~\cite{eisert2025mindgapsfraughtroad}. 
We are now in the \emph{noisy intermediate-scale quantum} (NISQ) era, characterized by quantum processors with hundreds to low-thousands of qubits that remain noisy and error-prone.
As of 2025, representative platforms include 56‑qubit trapped‑ion processors~\cite{liu2025certified, decross2025computational}, 127‑qubit superconducting processors~\cite{kim2023evidence}, and neutral‑atom processors reaching up to 280 qubits~\cite{bluvstein2024logical}.
One of IBM’s most advanced processors is Heron\footnote{\url{https://docs.quantum.ibm.com/guides/processor-types}}, with 156 qubits and no full error correction yet. IBM Quantum System Two, currently under development, connects multiple Heron processors and aims to scale to 4{,}000 qubits by the end of 2025 and 100{,}000 qubits in the 2030s. Meanwhile, the two‑qubit gate error rates approach 0.1\%~\cite{decross2025computational, graham2022multi}, and single‑qubit gate errors are below 0.01\%~\cite{li2023error, rower2024suppressing}.
% 0.003\%
Keeping pace with the rapid evolution of quantum computing, our goal is to identify the right applications and design a quantum database that delivers meaningful real‑world value.

\emph{Quantum data} is the information represented as quantum states. Such data may be classical data encoded into qubits, e.g., images or vectors embedded for quantum machine learning algorithms~\cite{biamonte2017quantum}; or  natively quantum outputs produced by quantum devices such as quantum sensors~\cite{huang2022quantum}. Consistent with classical database foundations~\cite{ramakrishnan2003database, elmasri2020fundamentals, garcia2008database, abiteboul1995foundations}, we define a \emph{quantum database} as an organized collection of quantum data. A \emph{quantum database management system (QDBMS)} is the software that manages quantum data and makes it useful. 
A \emph{quantum database system} is the end‑to‑end stack comprising the quantum database, the QDBMS, and the underlying quantum (and supporting classical) hardware. 
Quantum database systems employ quantum phenomena such as superposition, entanglement, and measurement to realize functionality that is inefficient or impossible for classical database systems. 
In the NISQ era, practical quantum database systems are most likely to be implemented as hybrid quantum–classical systems~\cite{hai2025quantum}.

Existing studies indicate three broad motivations of quantum database applications. The first is \textbf{database search.}   Grover's algorithm \cite{grover1996fast} can search unstructured data with complexity of $\mathcal{O}(\sqrt{N})$, where $N$ is the size of the domain. However, recent work indicates that 
NISQ devices cannot achieve the promised quadratic speedup \cite{chen2023complexity}. Moreover, classical databases have more mature implementations, such as B‑trees or hash indexes. Thus, the opportunity window for search acceleration is, for now, narrow.
The second possibility is using a quantum database to \textbf{store native quantum data}, such as the quantum states emitted directly by quantum sensors.  While compelling, these workloads are niche and do not motivate a general‑purpose quantum database.

\medskip
\para{Position} We explore another promising near‑term possibility, that is, employ quantum databases for \textbf{privacy}. 
Quantum cryptographic primitives such as  quantum key distribution \cite{BennettBrassard1984}, quantum private  queries \cite{giovannetti2008quantum}, and 
quantum oblivious transfer \cite{bennett1991practical}
% quantum bit commitment \cite{mayers1997unconditionally, lo1997quantum} 
have shown potential in regulated domains such as banking and finance \cite{pistoia2023paving, mas_qkd_sandbox_2025}, privacy-constrained healthcare \cite{kalaivani2021enhanced, tanizawa2025quantum} and biometric verification %(fingerprint matching) 
\cite{ramos2025securemultipartybiometricverification}. 
Building on these primitives, we investigate employing quantum databases for private information retrieval (PIR)  \cite{kushilevitz1997replication} and its stronger variant, symmetric PIR (SPIR) \cite{gertner1998protecting, di2000single}. PIR ensures user privacy: a user queries a record from the database server without revealing to the server which record it is retrieving. 
SPIR additionally enforces data privacy so the user learns only the requested tuple. PIR is already deployed in production. For example, Apple reports PIR‑backed visual search for photos ~\cite{apple2024hepir} 
%Mix of QKD and QOT on top of MPC
and live caller‑ID lookup~\cite{apple_pir_service_example_2025}. For SPIR, although not in production, recent experiments demonstrate feasibility over quantum‑secure networks for private biometric retrieval with formal server‑privacy guarantees~\cite{wang2022experimental}. 
We contend that privacy is an immediate, high‑impact opportunity for quantum databases.

\begin{figure}[tb]
    \centering
    \includegraphics[width=0.46\textwidth]{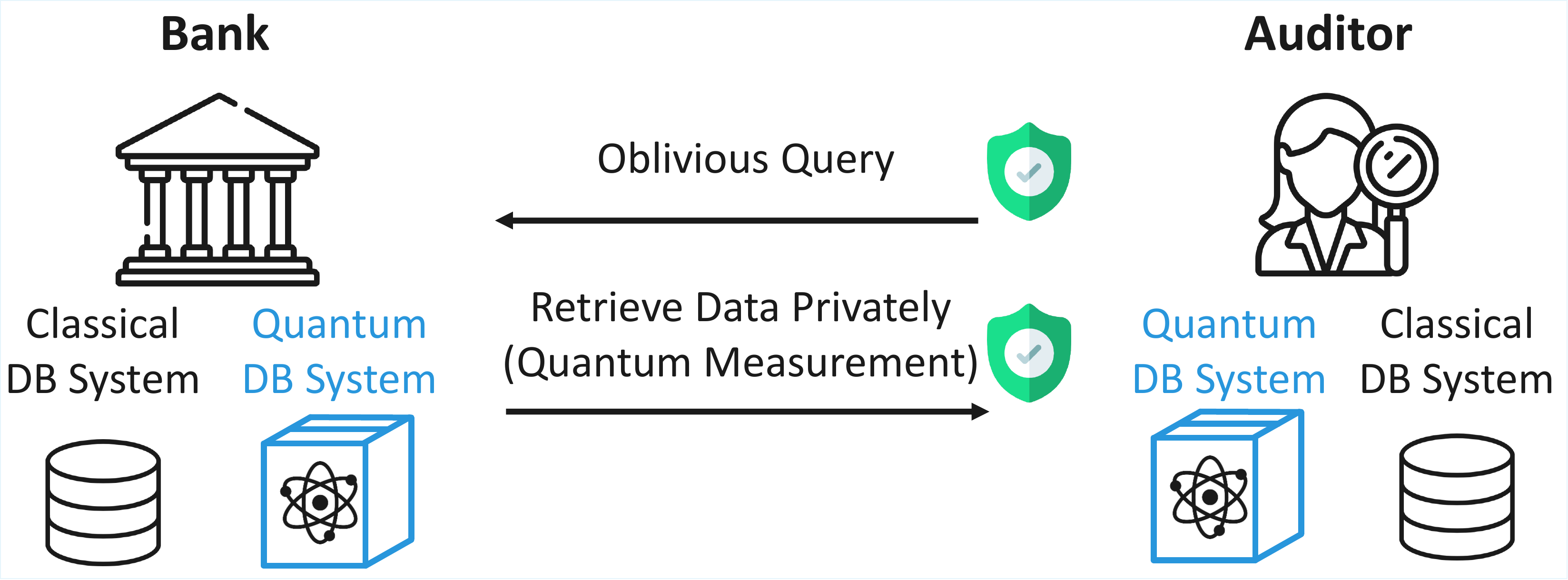}
    \caption{Example: Auditing with user and data privacy}
    \label{fig:example}
    \vspace{-0.4cm}
\end{figure}

\medskip
\para{Problem Statement} We target a \emph{private quantum database system} that enforces both \emph{user privacy} (the server learns nothing about the query) and \emph{data privacy} (the client learns only the requested record). The following example illustrates the setting.

\begin{myframe}[innerleftmargin = 5pt]
\label{exmpin:1}
\bemph{Example 1.}  
An auditor requires data about a customer from a bank’s databases while satisfying two privacy constraints: first, the auditor’s query (which customer is being audited) must remain oblivious to the bank; second, the auditor must learn nothing beyond that customer’s records. Ideally, quantum database systems, used alongside the classical database systems, facilitate privacy-preserving audits without exposing either the auditor’s intent or other clients’ data. \looseness=-1
\end{myframe}

\para{Contributions}
We propose a \emph{private quantum database} that realizes restricted queries for PIR/SPIR by encoding tuples as quantum states and leveraging two fundamental facts: (i) %projective 
measurement is destructive, so reading one basis erases information about the others; and (ii) the \emph{no-cloning theorem} \cite{wootters1982single} prevents duplicating unknown quantum states. Together, these physics-level constraints let a single server offer user privacy and bounded data privacy without trusted hardware or heavyweight cryptography. 

Concretely, our design: (1) maps keys to mutually unbiased bases and encodes blocks via quantum random access code, so that measuring the chosen basis reconstructs only the requested row while making other rows physically inaccessible; (2) quantifies the correctness–privacy trade-off through the copy budget $k$ and majority-vote amplification; and (3) introduces a hybrid quantum–classical architecture that stores classical descriptions of the prepared states inside a standard DBMS, regenerating fresh instances at query time, practical for NISQ hardware and compatible with today’s database systems.

This framework reframes “quantum advantage” for databases around two-sided privacy rather than speedups, and offers a concrete path to application-driven adoption of quantum components in future database systems.

\section{Preliminaries}
\label{sec:sol}
We first provide some quantum background required for understanding the paper. Readers seeking a more detailed exposition may refer to our previous work \cite{j7gt-zm6f,gatti2023random}.  

\subsection{Quantum Basics}
\label{ssec:pre}
A \emph{quantum bit (qubit)} is the analogue of a classical bit but can be in a \emph{superposition} of the computational basis\footnote{Column-vector notation:
$\ket{0}=\bigl[\,1\;0\,\bigr]^{\mathsf T}$ and
$\ket{1}=\bigl[\,0\;1\,\bigr]^{\mathsf T}$.}
states $\ket{0}$ and $\ket{1}$. That is, a general qubit is of the form
\[
\ket{\psi}=\alpha\ket{0}+\beta\ket{1},
\qquad
|\alpha|^2+|\beta|^2=1,
\]
with complex amplitudes $\alpha,\beta$.  
For $n$ qubits we use the \emph{tensor product}
$\ket{\psi_1}\otimes\cdots\otimes\ket{\psi_n}$, which expands as a linear combination of basis states such as $\ket{1}\otimes\ket{0}\otimes\ket{1}$; we typically drop $\otimes$ and write $\ket{101}$.  
% A state that cannot be written as a product of its subsystems is \emph{entangled}; e.g., the Bell pair $(\ket{00}+\ket{11})/\sqrt{2}$. 
In the gate-based quantum computation model, computation proceeds by applying unitary matrices $U$ ({\em quantum gates}) to these states.

\subsection{Quantum Measurement and Mutually Unbiased Bases}
Quantum mechanics forbids direct inspection of quantum states; instead, states (e.g., a computed $U\ket{\psi}$) can only be accessed via measurement. For $n$ qubits, we choose a \emph{measurement basis} $V=\{\,\ket{v_i}\,\}_{i=0}^{2^n-1}$, a complete set of $2^n$ orthonormal $n$-qubit states with $\langle v_i| v_j\rangle=\delta_{ij}$, where $\delta_{ij}=1$ if $i=j$ and $0$ otherwise. A projective measurement in $V$ collapses $\ket{\psi}$ to $\ket{v_i}$ with probability $p_i=\bigl|\braket{v_i|\psi}\bigr|^2$, and the ``measurement apparatus'' records the outcome $i$.

Two orthonormal measurement bases $V=\{\ket{v_i}\}$ and $W=\{\ket{w_j}\}$ on $n$ qubits (so $i,j\in\{0,\dots,2^n-1\}$) are \emph{mutually unbiased} if measuring a state prepared in one basis yields a \emph{uniform} outcome distribution in the other:
$\bigl|\braket{v_i|w_j}\bigr|^2=\frac{1}{2^n}$ for all $i,j$. An $n$-qubit system has \emph{dimension} (also called the number of \emph{levels}) $d=2^n$, and admits at most $d{+}1=2^n{+}1$ MUBs. 
More generally, a $d$-level system with $d$ a prime power admits at most $d{+}1$ MUBs (and such complete sets are known to exist)~\cite{wootters1989optimal}.

\subsection{Quantum Random Access Codes}
\label{ssec:qrac}
An $m\rightarrow n$ quantum random access code (QRAC) encodes $m$ classical bits into an $n$-qubit quantum system, with $m>n$. The goal is to maximize the probability of recovering any chosen bit by performing measurements on the quantum system. In high-dimensional settings ($d=2^n$), optimal QRACs exploit all $d+1$ MUBs available for prime-power dimensions by preparing a quantum state $\ket{\psi}$ maximizing the overlaps ($\braket{v_r|\psi}>1/d$) with the designated target basis states $\ket{v_r}$, one chosen in each MUB. This balances the measurement biases across all retrieval choices to an extent, and maximizes the average success probability.~\cite{ambainis2009quantumrandomaccesscodes}

\begin{figure}[tb]
    \centering
    \includegraphics[width=0.25\textwidth]{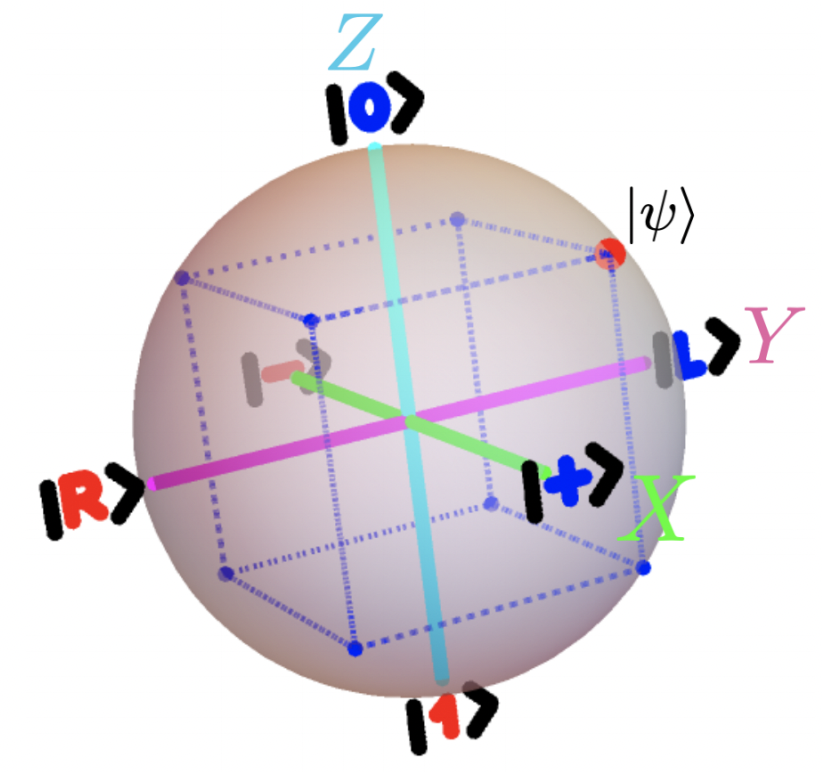}
    \vspace{-0.3cm}
    \caption{Bloch-sphere representation of $3\rightarrow 1$ QRAC, which encodes $3$ bits in a single qubit, and allows retrieval of any one of them with high probability. %In this representation, 
    All possible single-qubit quantum states lie on the surface of the sphere}
    \label{QRAC_3to1}
    \vspace{-0.5cm}
\end{figure}

As an example, consider a $3\rightarrow 1$ QRAC. For a single qubit, the three MUBs are the eigenbases of the \emph{Pauli} observables\footnote{The Pauli observables are $X$, $Y$, and $Z$, i.e., spin measurements along the Bloch $x,y,z$ axes, with matrices
$X=\begin{psmallmatrix}0&1\\[1pt]1&0\end{psmallmatrix}$,
$Y=\begin{psmallmatrix}0&-i\\[1pt]i&0\end{psmallmatrix}$,
$Z=\begin{psmallmatrix}1&0\\[1pt]0&-1\end{psmallmatrix}$.} $X$, $Y$, and $Z$. Consider encoding the bit string `$000$'. We choose a state $\ket{\psi}$ that lies midway (on the Bloch sphere; see Fig.~\ref{QRAC_3to1}) between the $+1$ eigenvectors of $X$, $Y$, and $Z$, that is, an intermediate state between $\ket{+}=\tfrac{1}{\sqrt{2}}(\ket{0}+\ket{1})$, $\ket{L}=\tfrac{1}{\sqrt{2}}(\ket{0}+i\ket{1})$, and $\ket{0}$. 
This can be visualized as one corner of a cube embedded within the Bloch sphere, shown as the red point (see Fig.~\ref{QRAC_3to1}). It exhibits a bias towards $\ket{+}$ when measured in the $X$ basis, towards $\ket{L}$ when measured in the $Y$ basis, and towards $\ket{0}$ when measured in the $Z$ basis. Specifically, these desired outcomes occur with probability
\[
P_{\mathrm{succ}}=\frac{1+1/\sqrt{3}}{2}\approx 0.789,
\]
the \emph{success probability} of the $3\rightarrow 1$ QRAC for any of the three retrieval choices.
Other codewords (e.g., $xyz$ with $x,y,z\in\{0,1\}$) correspond to flipping the Bloch-vector signs along the respective axes, yielding the other seven cube corners.

\section{Private and Restricted Database Queries}
We introduce \emph{quantum symmetric private information retrieval} (quantum SPIR), a cryptographic task in which a user retrieves a record from a server-held database while hiding the query (\emph{user privacy}, as in PIR~\cite{kushilevitz1997replication}) and, simultaneously, learning nothing beyond the requested record (\emph{data privacy}, as in SPIR~\cite{di2000single}). Following the standard in quantum information, we formalize our solution as a protocol that specifies the quantum–classical message flow and proves privacy and correctness probabilistically. 

In our setting, the parties (user and server) are quantum-capable and may exchange qubits, prepare quantum states, apply gates, and perform measurements over quantum channels~\cite{kerenidis2004quantum}. Conceptually, the classical database is transiently encoded into quantum states, with each data key mapped to a distinct measurement basis. The server prepares and sends a bounded number of quantum state copies\footnote{Preparing copies of {\em known} quantum states does not contradict the No‑Cloning Theorem~\cite{nielsen2010quantum}, which applies to unknown states.}, and the user measures in the basis corresponding to the desired key. Because quantum measurement is destructive and incompatible bases disturb the state, attempting to access many items reduces per-item success, creating an intrinsic trade-off between correctness and the amount of information extracted. This trade-off, together with the basis design, yields user privacy and enforces restricted access for data privacy. As such, the proposed {\em quantum database} is a classical database with quantum-enhanced query functionalities, providing full user privacy and setting an inverse proportionality between correctness and query size, which grants partial data privacy. We first provide a detailed example in Sec.~\ref{ssec:exampleqrac}; the general protocol is presented in Sec.~\ref{ssec:qdb}. Correctness and system considerations are discussed in Sec.~\ref{ssec:guaranteesystems} and we conclude in Sec.~\ref{ssec:simul} with simulation results.

 \subsection{Introductory Example}
 \label{ssec:exampleqrac}
We provide an example of our protocol featuring Alice (the server/database owner) and Bob (the client/querier).

\begin{myframe}[innerleftmargin = 5pt]
\label{exmp:2}
\bemph{Example 2}  
Alice has a database containing personal information of $5$ users:
\begin{center}
\begin{adjustbox}{width=\columnwidth,center}
\begin{tabularx}{1.22\columnwidth}{@{}l l X@{}}
\toprule
\textbf{Full Name} & \textbf{Phone} & \textbf{Email} \\
\midrule
Anton Brown & (+31) 123-4567 & \texttt{anton.brown@gmail.com} \\
Brian Smith   & (+42) 234-5678 & \texttt{bsmith@yahoo.com} \\
Cynthia Lee   & (+51) 345-6789 & \texttt{cynthia.lee@outlook.com} \\
Daniel Kim    & (+64) 456-7890 & \texttt{dkim@icloud.com} \\
Eva Martín  & (+34) 567-8901 & \texttt{eva.martin@aol.com} \\
\bottomrule
\end{tabularx}
\end{adjustbox}
\end{center}

For better understandability, we use the full name as the public identifier (key) in this example. 
Assume each phone number can be represented as a $30$-bit integer, and each email as $100$ bits by using $11$ characters of $8$-bit ASCII for the local part and $12$ bits to index one of $4096$ common domains. Thus, the private payload per user (phone+email) is $130$ bits. 
Here, we omit the full name from this count since the key is known to both parties and is not part of the protected payload.
% \fg{it is a bit strange that names are left out here}

Bob wishes to learn the information of exactly one user without Alice discovering which one, and Alice does not wish to reveal more of her database than necessary. For simplicity, we assume the values in the \emph{Full Name} column are unique and known to both Alice and Bob; they act as the lookup keys. 

We will employ systems of $n=2$ qubits, which admit $d+1=5$ mutually unbiased bases (MUBs) since $d=2^n=4$. Conveniently, this matches the number of database rows ($R=5$). Because each data row has $C=130$ bits, Alice prepares $65$ independently generated quantum states $\ket{\psi^{\ell}}$ (i.e., performs $65$ two-bit-per-MUB QRAC encodings), where $\ell=0,1,...,64$. The state with index $\ell$ encodes bit positions $2\ell$ and $2\ell{+}1$ for \emph{all} users. Once Bob reliably recovers $130$ bits for a single user, he can reconstruct that user’s payload. Alternatively, Bob could attempt to recover two users’ payloads, but then he must split his limited supply of state instances across two measurement bases, lowering his overall fidelity due to the destructive nature of quantum measurement.

\smallskip
To illustrate, consider the state $\ket{\psi^{18}}$, which encodes the last two bits of the \emph{first} character of each email’s ASCII code, namely the final two bits of \texttt{a}, \texttt{b}, \texttt{c}, \texttt{d}, \texttt{e}. These are $01$, $10$, $11$, $00$, and $01$, respectively. Alice uses the following $5$ MUBs for two qubits:

\begin{center}
\vspace*{0.5ex}\begin{adjustbox}{width=1\columnwidth,center}
\begin{tabular}{|c|c|l|}
\hline
\textbf{MUB \#} & \textbf{Commuting observables} & \textbf{Description} \\
\hline
1 & $\{ Z \otimes I,\ I \otimes Z \}$ & Computational (product) basis \\
2 & $\{ X \otimes I,\ I \otimes X \}$ & $X$ basis on both qubits \\
3 & $\{ Y \otimes I,\ I \otimes Y \}$ & $Y$ basis on both qubits \\
4 & $\{ X \otimes Y,\ Y \otimes Z \}$ & Entangled basis \\
5 & $\{ Y \otimes X,\ Z \otimes Y \}$ & Entangled basis \\
\hline
\end{tabular}
\end{adjustbox}
\end{center}

Each row specifies one measurement Bob can perform, corresponding to one of the five users he may choose to read. Each such measurement has $4$ outcomes $00,01,10,11$, each associated with a basis state. To encode the target two-bit strings $(01,10,11,00,01)$ across the five bases, Alice prepares a single \emph{intermediate} state whose Bloch representation is equidistantly biased toward the desired outcome state in each MUB (the standard QRAC construction). She then sends $k$ independently prepared copies of this state, denoted $\ket{\psi^{18}}$, to Bob.

\smallskip
Bob chooses MUB \#1 and measures all $k$ copies in that basis. From known $2\!\rightarrow\!1$ QRAC performance~\cite{Casaccino_2008}, each copy yields the correct two-bit outcome (here, $01$) with average probability $0.5424$, and each of the other three outcomes with average probability $0.1525$. With $k$ i.i.d.\ trials, the probability that majority vote returns the correct outcome follows the corresponding multinomial distribution.

\smallskip
In the table below, we report the average success probability $p(M)$ to correctly reconstruct \emph{one} full payload ($130$ bits) when attempting to read $M\in\{1,2\}$ users in total, as a function of the number $k$ of copies per encoded two-bit block:
\[
\begin{array}{c|cc|c|cc}
\hline
k & p(1) & p(2) & k & p(1) & p(2) \\
\hline
1   & \sim  10^{-18} & \sim  10^{-18} & 41  & 0.8927 & 0.1524 \\
11  & 0.0006 & \sim  10^{-8} & 51  & 0.9709 & 0.3904 \\
21  & 0.1724 & 0.0004 & 61  & 0.9923 & 0.6227 \\
31  & 0.6429 & 0.0222 & 71  & 0.9980 & 0.7869 \\
\hline
\end{array}
\]

Notably, certain values of $k$ (e.g., $k=41$) yield a strong gap between the success probabilities for retrieving one vs.\ two users, enhancing data privacy: it is difficult for Bob to recover more than a single payload reliably. To quantify this, define the expected number of successfully retrieved rows $E_r(M)=M\,p(M)$ and the \emph{data-privacy index} $\mathcal{P}_i=E_r(i)/E_r(i+1)$. When $\mathcal{P}_i>1$, attempting $i{+}1$ rows produces \emph{less} total information than attempting $i$ rows.

\begin{adjustbox}{width=1.08\columnwidth,center}
\begin{minipage}{1.21\columnwidth}
\[
\begin{array}{c|ccc|c|ccc}
\hline
k & E_r(1) & E_r(2) &\mathcal{P}_1 & k & E_r(1) & E_r(2) &\mathcal{P}_1 \\
\hline
1   & \sim  10^{-18}  & \sim  10^{-18} & 1 & 41  & 0.8927 & 0.3048 & 2.93 \\
11  & 0.0006 & \sim  10^{-8} & \sim  10^4 & 51  & 0.9709 & 0.7808 & 1.24 \\
21  & 0.1724 & 0.0008 & 229.5 & 61  & 0.9923 & 1.2455 & 0.80 \\
31  & 0.6429 & 0.0444 & 14.49 & 71  & 0.9980 & 1.5738 & 0.63 \\
\hline
\end{array}
\]
\end{minipage}
\end{adjustbox}
\vspace*{0.2ex}

To construct a protocol where Bob is strongly incentivized to query at most $i$ rows, one can choose the largest $k$ with $\mathcal{P}_i>1$ (or $\mathcal{P}_i>2$ for stronger privacy). In this example, this occurs around $k=51$ (or $k=41$). To incentivize exactly $i>1$ rows, choose $k$ such that $\mathcal{P}_i>1$ but $\mathcal{P}_{i-1}<1$; in that case, computing $p(3)$ would be required.

\end{myframe}

% Our protocol guarantees:\fg{the following is moved up a bit. Would be great if it could be made more clear how quantum effects kick in (destructive nature, ...)}
% \begin{description}
%     \item[User privacy:] Simply because the use does not communicate with the server.
%     \item[Data privacy:] The client performs a query restricted by the number of state copies $k$ sent by server. If the client tries to query additional database rows, the fidelity of the retrieved records drops, and potentially becomes unusable.
%     \item[Correctness:] The client is able to retrieve the desired item with high enough probably.
% \end{description}
% We consider a single-server, information-theoretic SPIR setting. \gc{can you please check the previous sentence?}

\subsection{The General Protocol}
\label{ssec:qdb}
% A high-level illustration of our proposed approach is depicted in Fig.~\ref{fig:sys}. 
We now present our protocol in full generality.
\begin{enumerate}[leftmargin=1em]
\item \textbf{Setup (keys $\leftrightarrow$ MUBs).}
Alice holds a classical database of size $R\times C$, where $R$ is the number of rows (keys) and $C$ is the number of \emph{binary columns} (bits) per row to be protected. Bob only knows the public keys (e.g., unique names). Alice and Bob agree in advance on a bijection between the $R$ keys and a list of $n$-qubit MUBs, so that each key/row is associated with one MUB. This maps each classical row to \emph{one} quantum measurement basis, avoiding redundant encodings. The number of rows must satisfy $R \le 2^n+1$ to ensure enough MUBs (since $d=2^n$ admits at most $d{+}1$ MUBs).

\item \textbf{Blockwise QRAC encoding.}
Partition each row’s $C$ bits into $L=\lceil C/n\rceil$ contiguous blocks of $n$ bits. For block index $\ell\in\{0,\dots,L-1\}$, let $b_{r,\ell}\in\{0,1\}^n$ denote the $n$-bit block of row $r$. In the MUB assigned to row $r$, let $\ket{v^{(r)}_{b_{r,\ell}}}$ be the basis state labeled by $b_{r,\ell}$. 
We \emph{encode} block $\ell$ by constructing an   
% \fg{what is pure in ICDE terms?}
$n$-qubit state $\ket{\psi^{\ell}}$\footnote{Here we mean pure states.  A pure state is a quantum state with complete information. It can be described by a single state vector rather than a mixture of several possible states. No classical randomness is involved.} that is (QRAC-) biased\footnote{An equidistant bias is desirable for our protocol, but not optimal for QRACs of more than one qubit.} toward the $R$ target states $\{\ket{v^{(r)}_{b_{r,\ell}}}\}_{r=1}^R$, i.e., it equalizes (and maximizes) the success probability across the $R$ chosen MUBs for that block.
Alice stores a \emph{classical description} of how to prepare each $\ket{\psi^{\ell}}$ (e.g., circuit parameters); no quantum storage is required.
% \gc{Footnote on bias equidistance added. Could be important when actually doing the implementation}

\item \textbf{Private query request and rate limit $k$.}
Bob requests a restricted, private query. Alice grants the request and sets a per-block sending budget of $k$ \emph{independently prepared} instances of each state $\ket{\psi^{\ell}}$.
This does \emph{not} invoke cloning of an unknown state; Alice prepares fresh instances from the known classical description. 
\item \textbf{Transmission.}
For each block $\ell$, Alice prepares and sends $k$ instances of $\ket{\psi^{\ell}}$ to Bob over a quantum channel and, over a classical channel, communicates the block index $\ell$ (only). Sending $\ell$  reveals no row information and is required so Bob can group the received states by block. 
Alice must make use of an appropriate quantum hardware and quantum circuit setup. (Implementation details of the state preparation and hardware are discussed in Sec.~\ref{sec:hybrid}.) 

\item \textbf{Bob’s measurement choices.}
Bob selects $M\in\{1,\dots,R\}$ MUBs (thus $M$ keys/rows) to attempt to read. For each block $\ell$, he divides the $k$ received instances among the $M$ chosen MUBs (ideally using an odd number per MUB for majority voting) and measures each instance in the corresponding basis, recording the $n$-bit outcomes.

\item \textbf{Reconstruction.}
For each chosen row and each block $\ell$, Bob applies majority vote over the outcomes measured in that row’s MUB to produce an $n$-bit estimate $\hat b_{r,\ell}$. Concatenating the blocks yields an estimated row $\hat x_r\in\{0,1\}^C$. He does so with success probability inversely dependent on $M$: the number of database keys he attempted to query.
% \fg{what is fidelity actually? Not sure I understand this.}\gc{Meant success probability. Will revert to that.}\fg{:-) Every discipline has some many concepts .. best to go easy on the db folks and keep things to minimal}
\end{enumerate}

\subsection{Protocol Guarantees \& Systems Considerations}
\label{ssec:guaranteesystems}
We next quantify the success-probability of the quantum query protocol, make explicit the prior agreements between Alice and Bob, and
discuss practical feasibility and scalability.

\para{Success probability} Given a QRAC that uses an $n$-qubit state to encode $n$ bits in each of up to $2^n+1$ MUBs, the average success probability of retrieving the $n$ bits correctly can be estimated. This number is known for small QRACs, such as the $n=1,2,3$ QRACs, which have respective success probabilities $p=0.789,0.5424,0.3372$~\cite{Casaccino_2008}. Note that this probability can be as low as $1/2^n$, since it attempts to guess $n$ bits. This success probability is amplified by means of majority voting of $k$ quantum-state copies, which we estimate with multinomial distributions. We simplify these distributions by assuming that all unsuccessful outcomes have the same probability, such that the  probability of obtaining $k_j$ counts for each bitstring configuration $j=1,2,...,2^n$ when measuring the quantum state $k$ times, where $k=\sum_j k_j$,  is given by
\begin{equation}
\!\!P(k_1, k_2, \ldots, k_{2^n}) \!=\!
\frac{N!}{k_1!\, k_2! \cdots k_{2^n}!}
\,
p^{k_1}
\!\!\left( \frac{1 - p}{2^n - 1} \right)^{N-k_1}.\end{equation}

We employ this expression to approximate the probability that the correct outcome is the one with most counts. This is the process we refer to as {\em majority voting}.

\medskip
\para{Prior agreements}
Before any query is executed, Alice (data owner) and Bob (client) agree on:
\begin{itemize}
\item The database shape $R\times C$. This shape enforces the use of $R$-MUB encodings in $n$-qubit, using the smallest $n$ such that $R\leq 2^n+1$. This shape also defines the number of required encodings to be $\lceil C/n \rceil$.
\item 
a bijection between the $R$ keys and a list of $n$-qubit mutually unbiased bases (MUBs), with ordered observables. Ordering the observables inside of each MUB is necessary for measurements to yield bitstrings. 
% \gc{THIS ordering is necessary. Each MUB must have ordered observables, so that they do produce a bitstring when measured}
% A bije on the $R$ MUBs, so as to map each one to a database key (row).\fg{so the db is required to consist of ordered rows as well?}
%\item The MUB observables of $n$ qubits to use  ordering for them, so as to associate MUB measurement results to ordered bitstrings. 
% \fg{it would be helpful if this could be tied to the steps in the protocol. Not sure I understand this.}
\item Optionally, the budget of quantum state copies $k$ to send. This number is decided by Alice, but it is convenient to communicate it to Bob, so that he can make an informed decision on what query size and correctness to aim for by planning an appropriate measurement scheme.
\end{itemize}
Note that these prior agreements only depend on the $R\times C$ database shape. Specifically, they do not depend on the database contents, nor do they depend on Bob's query choices. 
% \fg{Should we add: and in particular, they don't depend on what Bob will be asking right?!}\gc{Added a small phrase on that. Give it a look (is the "Bob's" too informal?)}\fg{good sentence. And, right level of formality, no worries}\gc{ok, great}\gc{loved the changes to the section}\fg{it was good already, so added minor things.}\gc{let's jump to III-B for a moment. there's an important section i want to review}\fg{ok.}

\medskip
\para{Technological feasibility} We argue that the protocol is within reach in practice.
\begin{itemize}
\item Quantum enhanced querying is not focused on providing a speedup, but rather two-way privacy, so it does not require quick loading of the classical information into quantum states, a common caveat of most quantum algorithms.\looseness=-1
\item Random noise in the preparation and measurement of quantum states only changes the baseline QRAC success probability, which simply shifts the required number of copies $k$ for a given correctness and query size. That is, random decoherence in the quantum channel can be corrected by simply increasing a parameter in the protocol. \looseness=-1
\item A proof of concept of the quantum database query can be performed in superconducting circuits, but a real implementation requires a quantum channel between two parties, which is better suited to single-photon setups.
\item In terms of scalability,  encoding a modern square database of 
% Terabyte size (
$3$-million rows only requires to prepare and measure $22$-qubit quantum states, which is feasible within the limits of current quantum technology. The current challenge to implement a quantum query of this scale lies in computing the states of $22$-qubit QRACs (employing all available MUBs), which we reckon should be possible, but demands a dedicated future line of work.
\end{itemize}

\begin{table}[t]
\caption{Preliminary simulation results}
\centering
\begin{tabular}{|c|c|c|}
\hline
\textbf{MUB \#} & \textbf{Expected values} & \textbf{Success probability $p$} \\
\hline
1 & $+0.620,\,+0.628$ & $0.6592$ \\
2 & $-0.299,\,-0.307$ & $0.4244$ \\
3 & $-0.365,\,-0.345$ & $0.4591$ \\
4 & $-0.330,\,-0.630$ & $0.5420$ \\
5 & $-0.308,\,-0.607$ & $0.5255$ \\
\hline
\end{tabular}
\label{tab:preliminary}
\end{table}
 
\subsection{Preliminary Simulation Results}
\label{ssec:simul}
% \gc{Here's a simulation I did for the 2q case, using brute force phase-space search in Mathematica}
% To validate feasibility and calibrate the copy budget \(k\) used for majority‑vote amplification, we evaluate a minimal instance of our protocol. 
As a proof of concept, we have conducted an experiment to validate our proposed quantum protocol. 
We have implemented the protocol for a simple 2-qubit QRAC case and simulated it using Wolfram Mathematica\footnote{\url{https://www.wolfram.com/mathematica/}}. The code is available online\footnote{\url{https://doi.org/10.5281/zenodo.17568665}}, and Table~\ref{tab:preliminary} reports the results.

\medskip
\para{Implementation and experimental setup}
We have encoded a $5\times 2$ binary ($+1$s or $-1$s) table using $2$-qubit quantum states, choosing MUB $1$ to correspond to observable values $+1,+1$ and choosing MUBs $2,3,4,5$ to correspond to observable values $-1,-1$, following the order of the table in the previous example.
% \fg{this table could perhaps be omitted since it was already given in example, but perhaps just say mapping mub\# to Bits?}
% \gc{Agreed. Its sufficient to say that MUB 1 corresponds to bits +1,+1 and the rest to bits -1,-1. I'll comment the table, just in case}\fg{Hi Giancarlo! You are night owl? Or at a different time zone? Night owl, but have to wake at 5:30am}\fg{omg, if you have energy left, could you read through  sec III and removed comments. Left only a few. }
% \fg{Thanks. I also switched order, example first, then general protocol}\fg{I will check IV now}
%\begin{center}
%\begin{tabular}{|c|c|c|}
%\hline
%\textbf{MUB \#} & \textbf{Observables} & \textbf{Bits} \\
%\hline
%1 & $\{ Z \otimes I,\ I \otimes Z \}$ & $+1$,\,$+1$ \\
%2 & $\{ X \otimes I,\ I \otimes X \}$ & $-1$,\,$-1$ \\
%3 & $\{ Y \otimes I,\ I \otimes Y \}$ & $-1$,\,$-1$ \\
%4 & $\{ X \otimes Y,\ Y \otimes Z \}$ & $-1$,\,$-1$ \\
%5 & $\{ Y \otimes X,\ Z \otimes Y \}$ & $-1$,\,$-1$ \\
%\hline
%\end{tabular}
%\end{center}
We numerically computed complex amplitudes $a_i$ for a quantum state of the form $\ket{\psi}=a_1 \ket{00}+a_2 \ket{01}+a_3 \ket{10}+a_4\ket{11}$ so as to make the $Z \otimes I$ and $I \otimes Z$ expected values as close to $+1$ as possible, and make the rest as close as possible to $-1$, via least squares optimization. This simulation was performed with stochastic Hilbert-space search in Mathematica, maximizing the desired observable expected values and decreasing the search space at each iteration.

\medskip
\para{Result analysis}
Up to a global phase, we obtain \(a_1=0.8162\), \(a_2=-0.2424-0.2917i\), \(a_3=-0.2340-0.3047i\), \(a_4=0.0645-0.1955i\), yielding the expected values and success probabilities in Table~\ref{tab:preliminary}. For a Pauli observable with expectation \(E\), the probability of outcome \(+1\) is \((E+1)/2\); we compute the per‑MUB success by multiplying the two corresponding probabilities. The mean success across MUBs is \(0.5220\), consistent with known 2‑qubit QRAC behavior~\cite{Casaccino_2008}. Shared randomness between Alice and Bob can equalize row‑wise success by randomizing key–basis assignments.

 The simulation confirms that a 2‑qubit QRAC/MUB encoding enables the retrieval with an average success of approximately  $0.52$ (best‑case approximately $0.66$), aligning with previous results and guiding the choice of copy budget \(k\) for majority‑vote amplification in our quantum database.

\section{Hybrid Approach}\label{sec:hybrid}
In this section, we show  how to integrate the quantum database consisting of quantum states encoding data with classical relational databases.

 \subsection{Quantum States in RDBMS}
\label{ssec:qdata_rdbms}
% The previously introduced protocol may generate large number of intermediate quantum states \rh{how many?}, which we can store in relational database. This will save the expensive quantum resources and make our approach more suitable for NISQ devices that have limited qubits. 
% We leverage  relational databases to store metadata such as circuit parameters for preparing qubit states in our proposed quantum protocol. Alternatively, one can precompute qubit states and store their classical representations directly in the relational databases. 

In our quantum SPIR protocol (Sec.~\ref{ssec:qdb}), Alice must send, for each block
$\ell\in\{0,\dots,L-1\}$ with $L=\lceil C/n\rceil$, up to $k$ independently prepared
instances of a known pure state $\ket{\psi^{\ell}}$ to Bob. On NISQ hardware,
\emph{recompiling} and \emph{recalibrating} the state-preparation circuit for every
run is costly. Instead, Alice precomputes a \emph{classical description} of each
state (or of its preparation circuit) and stores it in a relational DBMS. At
query time, she (re)generates fresh instances from this description. To this aim, we need to store, for each block $\ell$:
(i) the mapping from database key/row $r$ to its MUB index (fixed, see Sec.~\ref{ssec:qdb});
(ii) the per-row target basis label $b_{r,\ell}\in\{0,1\}^n$ (the $n$ bits of row $r$
that block $\ell$ encodes);
(iii) a classical description of the \emph{intermediate} QRAC state $\ket{\psi^\ell}$
(or of a circuit $U_\ell$ such that $U_\ell\ket{0}^{\otimes n}=\ket{\psi^\ell}$).
Items (i)–(iii) are classical and fit naturally in tables; they let Alice recreate
$\ket{\psi^\ell}$ on demand and dispatch $k$ fresh instances when Bob initiates a
query.

We consider standard classical representations of quantum states used in simulation
and compilation pipelines: state vectors~\cite{nielsen2010quantum},
tensor networks~\cite{biamonte2020lecturesquantumtensornetworks},
decision diagrams~\cite{wille2022decision}, and the stabilizer
tableau~\cite{aaronson2004improved}. Our existing system,
Qymera~\cite{littau2025qymera}, already supports state vectors, tensor networks, and
decision diagrams in relational engines such as SQLite and DuckDB. For
completeness and because several ingredients of our protocol 
% (Pauli/MUB measurements, Clifford pre- and post-processing, authentication) 
(e.g., measurements, MUB observables) 
are stabilizer-heavy,
we also support the \emph{tableau} representation inside an RDBMS.

\subsection{Tableau Representation}\label{ssec:tablleau}
The $n$-qubit Pauli group is
$\mathcal P_n=\{\pm P_1\otimes\cdots\otimes P_n\mid P_j\in\{I,X,Y,Z\}\}$ with $I$ identity and $X$, $Y$ and $Z$ the Pauli observables encountered earlier.
A pure state $\ket{\psi}$ is a \emph{stabilizer state} if there exists a set of
$n$ commuting, independent Paulis $\{g_1,\dots,g_n\}\subset\mathcal P_n$ such that
$g_i\ket{\psi}=\ket{\psi}$ for all $i$. These $n$ generators generate a group of
size $2^n$ that fixes $\ket{\psi}$ and only $\ket{\psi}$.
For example, on two qubits, $\ket{00}$ is stabilized by the \emph{group}
$\{II,\,ZI,\,IZ,\,ZZ\}$, but only \emph{two} commuting generators, say $ZI$ and
$IZ$, are \emph{independent}; $ZZ=(ZI)\,(IZ)$ and $II$ is the identity. Hence two
generators suffice for $n{=}2$. The de facto compact encoding for stabilizer states is the \emph{tableau}
\cite{gottesman1997stabilizer, aaronson2004improved}. For $n$ qubits, the tableau
stores $n$ commuting generators as a triple $T=(X_M,Z_M,S_M)$ with
$X_M,Z_M\in\{0,1\}^{n\times n}$ and $S_M\in\{0,1\}^n$:
\begin{itemize}[leftmargin=*]
\item Row $i$ encodes generator $g_i$; column $j$ encodes the Pauli on qubit $j$.
      The pair $(X_M[i,j],Z_M[i,j])$ decodes to $I,X,Z,Y$ as $(0,0),(1,0),(0,1),(1,1)$.
\item The sign bit $S_M[i]=0$ ($+$) or $1$ ($-$) records the overall phase of $g_i$.
\end{itemize}
\begin{myframe}[innerleftmargin=5pt]
\label{exmp:1}
\bemph{Example 3 (tableau for two generators).}
Generators of the  stabilizer group of $\ket{00}$ are encoded as
\[
T=\left[
\begin{array}{cc|cc|c}
\multicolumn{2}{c|}{X_M} & \multicolumn{2}{c|}{Z_M} & S_M  \\\hline
0 & 0 & 1 & 0 & 1  \\
0 & 0 & 0 & 1 & 0 
\end{array}
\right],
\]
where indeed the first row encodes $-ZI$, the second $IZ$.
% Here the two rows are independent and commute; the full stabilizer group
% $\{II,ZI,IZ,ZZ\}$ is generated by them.
\end{myframe}
Stabilizers can be stored and manipulated \emph{compactly} as tableaux and updated by
bitwise row operations in $O(n)$ time, ideal for an RDBMS backend. When a block
state $\ket{\psi^{\ell}}$ is non-stabilizer (typical for QRAC intermediates),
we store its \emph{circuit parameters}, to regenerate it later on hardware.\looseness=-1

We record every Pauli symbol with two Boolean \emph{flags}:
the first flag stands for the presence of~$X$, the second for~$Z$
(Tab.~\ref{tab:flag-coding}).  Concatenating the $2n$ flags yields a
fixed‑width bit string that is compact yet easy to manipulate inside an
SQL engine. \begin{table}[h]
   \vspace{-0.2cm}
   \caption{Two‑flag coding of a single‑qubit Pauli operator}
\label{tab:flag-coding}
\centering
\begin{tabular}{c|cccc}
\textbf{Flags} $(x,z)$ & $(0,0)$ & $(1,0)$ & $(0,1)$ & $(1,1)$\\\hline
Pauli operator         & $I$     & $X$     & $Z$     & $Y$
\end{tabular}

    \vspace{-0.3cm}
\end{table}
Below we illustrate one possible schema definition using PostgreSQL syntax; other relational engines that support a fixed-length \texttt{BIT} datatype can be used in a similar fashion. Based on this schema, standard SQL queries can be written to retrieve and manipulate qubit-related information.

\begin{lstlisting}[
  language=SQL, 
  label=lst:schema, 
  caption={Schema for storing Pauli generators.},
  basicstyle=\ttfamily\scriptsize,
  breaklines=true,
  columns=flexible
]
CREATE TABLE generators (
  circ_id BIGINT,      -- circuit ID
  g_idx   SMALLINT,    -- generator index
  n       SMALLINT,    -- number of qubits
  flags   VARBIT NOT NULL, -- 2n bits = Pauli string
  neg     BOOLEAN,     -- TRUE = overall minus sign
  PRIMARY KEY (circ_id, g_idx),
  CHECK (bit_length(flags) = 2 * n)
);
\end{lstlisting}

\begin{figure*}[t]
    \centering
    \includegraphics[width=0.8\textwidth]{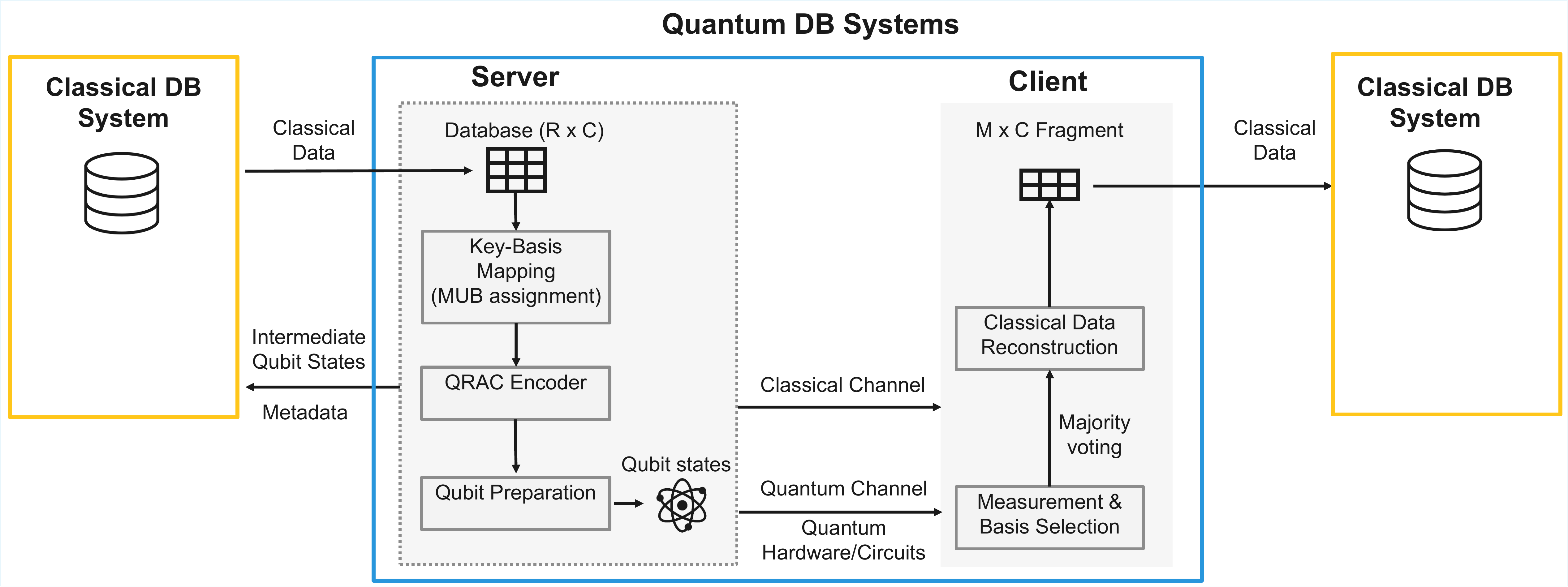}
     % \vspace{-0.2cm}
    \caption{Overview of our hybrid quantum-classical  approach  }
           % Alice encodes each row of her database into mutually unbiased
           % basis (MUB) states, sends $k$ authenticated copies per state to Bob,
           % who measures up to $M$ bases and, via majority voting, reconstructs an
           % $M\times C$ fragment — without revealing \emph{which} fragment he chose}
           
    \label{fig:sys}
    \vspace{-0.3cm}
\end{figure*}

\smallskip
\noindent

\subsection{Hybrid Quantum-Classical Approach}
Fig.~\ref{fig:sys} illustrates our hybrid approach. Classical data stored in a classical database system (e.g., an RDBMS) serves as the input to the quantum database system.  
On the \emph{server} side (Alice), the system consists of three components: (i) given a classical database with size of $R\times C$, \emph{Key–Basis Mapping}, which assigns each primary key to a mutually unbiased basis; (ii) the \emph{Quantum Encoder (QRAC)}, which transforms the classical data into \(n\)-qubit states; and (iii) \emph{Qubit Preparation}, which instantiates these states on quantum hardware and transmits \(k\) copies to the client.

On the \emph{client} side (Bob), retrieving the requested data involves two components. The \emph{Measurement \& Basis Selection} component chooses up to \(M\) measurement bases and performs measurements on the received qubits. The resulting outcomes are locally  processed via majority voting and reconstructed by the \emph{Classical Data Reconstruction} component into an \(M \times C\) fragment of the classical database, 
which can then be stored in a classical DBMS. \looseness=-1

The role of classical database systems (e.g., RDBMSs) includes storing the server- and client-side classical data, the classical descriptions of the intermediate QRAC states from which the transmitted qubits are prepared, together with the metadata that drives our protocol, namely the mapping from each key/row \( r \) to its assigned MUB index and the per-row target basis labels.

\subsection{Ongoing Implementation}
\label{sec:disc}
\begin{figure}[tb]
    \centering
    \includegraphics[width=0.47\textwidth]{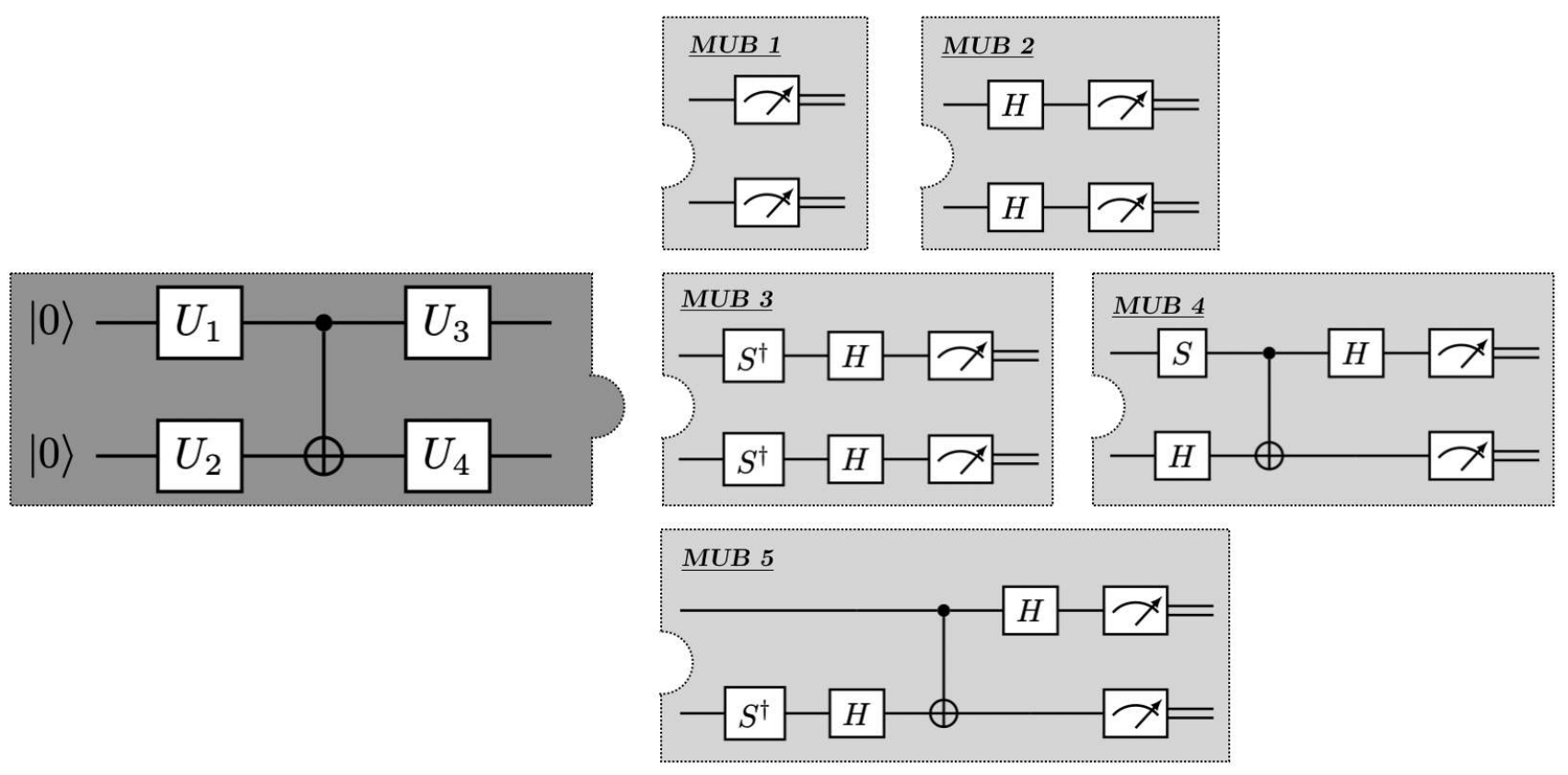}
    \vspace{-0.3cm}
    \caption{Quantum circuits for Alice to prepare any $2$-qubit state (dark grey) and for Bob to measure any $2$-qubit MUB (light grey). Each horizontal line represents a qubit that is transformed with quantum gates. The boxed $U_i$, $S$, $S^\dagger$ and $H$ are single-qubit gates typically implemented in photonics via waveplates, and the dot-line-circle symbol is a Controlled-NOT gate, typically implemented in photonics via interferometer setups.}
    \label{q_circuits}
    \vspace{-0.5cm}
\end{figure}

% \fg{this section may perhaps also be absorbed in previous one?}
We consider our example scenario (Sec.~\ref{ssec:exampleqrac}) as a starting point.
The quantum states sent by Alice in our database query protocol can be generated with the general quantum circuit in Fig.~\ref{q_circuits}, capable of preparing any $2$-qubit quantum state~\cite{nielsen2010quantum}. We also include corresponding quantum circuits for Bob to measure in any of the $5$ MUBs of $2$ qubits. Note that Alice prepares intermediate states between MUB elements, so there are $4^5=1024$ states to choose from ($4$ options per MUB), whereas Bob only requires to choose one of $5$ MUBs to perform measurements.

Superconducting circuits via cloud quantum computing are an appropriate hardware to test the quantum-state generation and measurement in our protocol. In particular, we plan to conduct experiments on the superconducting processors \textit{Tuna-5} and \textit{Starmon-7}, available through the Quantum Inspire infrastructure developed at TU Delft\footnote{\url{https://www.quantum-inspire.com/backends/transmons/}}. Moreover,  to implement a working proof of principle where Alice and Bob are in different locations, photonic systems are a more natural fit due to their suitability for long-distance quantum communication.   
% Superconducting circuits via cloud quantum computing are an appropriate hardware to test the quantum-state generation and measurement in the protocol, but photonic systems are better suited to implement a working proof of principle where Alice and Bob are in different locations. 
To this end, we can either employ two-photon systems, encoding each of the two qubits of the protocol in their polarization degree of freedom, or we can alternatively employ single-photon systems, encoding one qubit in the polarization degree of freedom and the other in the spatial-mode degree of freedom, as done in~\cite{Luda_2014}.

Moreover, we are extending our system Qymera \cite{littau2025qymera}, which stores the data representations of quantum states as relational tables and compiles quantum gate operations and measurement computations into SQL queries executed on RDBMS engines such as SQLite and DuckDB. 
% Qymera currently supports commonly used qubit state representations.
% , including the state vector and tensor network formalisms (e.g., matrix product states \cite{perez2007matrix, biamonte2020lecturesquantumtensornetworks}), de-
% cision diagrams,. 
As introduced in Sec.~\ref{ssec:tablleau}, we are now adding support for the tableau representation for stabilizer states. In parallel, we are developing a quantum database optimizer that (i) selects appropriate quantum states, (ii) tunes performance-related parameters such as $M$ and $k$, and (iii) balances workloads between quantum and classical execution. 
% For example, if quantum hardware is overloaded or decoherence rates exceed a user-defined threshold, the optimizer gracefully falls back to a purely classical execution plan.

% \section{BACKGROUND and PROBLEM STATEMENT}

\section{Related Work}
\label{sec:rw}
% \rh{Rihan will rewrite this:
% \\PIR applications (?) (one server for the price of two?)
% \\-Most QPQ are currently based on QKD. What about the one that isn't based on QKD: does it have the usual vulnerability to delayed-measurement attack?
% \\Make chapter 2.2 tell the story that we're treating a different problem: we don't require speedup
% }

This work lies at the intersection of data management, quantum computing, and privacy and security. To position our contribution, this section reviews prior quantum-related efforts within the database community (Sec.~\ref{ssec:landscape}--\ref{ssec:groverDB}), surveys existing work on classical PIR, SPIR, and their quantum variant QSPIR (Sec.~\ref{ssec:pqd}--\ref{ssec:qspir}), and relates our approach to other privacy and security techniques (Sec.~\ref{ssec:others}).

\subsection{Classical DB \& Quantum Computing}
\label{ssec:landscape}
% Recent works in the intersection of data management and quantum computing 
Recent quantum‑focused efforts within the DB community can be divided into two directions: first, 
leveraging quantum computers as new hardware to boost classical database tasks, such as query optimization \cite{trummer2016multiple, schonberger2023SIGMOD, schonberger2023quantumb}, data integration \cite{10.1145/3725226}, index selection \cite{10.14778/3681954.3682025}, and transaction management \cite{ groppe2021optimizing}. 
The other direction is to use database technologies for quantum computing \cite{hai2025quantum}, such as utilizing RDBMSs for running quantum circuit   simulation \cite{einsteinSQL2023,Trummer24, littau2025qymera}. 

Our work aligns primarily with the first research direction: we leverage quantum computing techniques (i.e., QRAC, MUBs) to design a quantum-native database system that supports symmetric private information retrieval. Meanwhile,  our implementation engages with the second research direction: Sec.~\ref{sec:hybrid} explains how we manage classical representations of quantum states within an RDBMS, building on our earlier efforts that use relational databases for quantum computing.

% A recent vision divided the landscape into three paradigms. 
% Althouh a quantum-native solution (introduced in Sec. 3.2) is suffieicent. We show that the hybrid solution of combining classical database brings a more efficient, and decohrece-resislent solution.

\subsection{Quantum DBMSs Based on Grover’s Algorithm}
\label{ssec:groverDB}

A \emph{database} is a persistent, electronically stored collection of data~\cite{abiteboul1995foundations,ramakrishnan2003database}.  
In the quantum community, the term \emph{database}\footnote{In \cite{roy2013quantum}, the authors use the term “quantum database” purely as a metaphor in a classical resource-allocation context. Their work is unrelated to quantum computing, and  falls outside the scope of this paper.} usually denotes a conceptual framework for processing and searching data using quantum algorithms. 
Grover’s algorithm searches an unsorted set of \(N\) items in \(\mathcal{O}(\sqrt{N})\) iterations instead of the classical \(\mathcal{O}(N)\) bound~\cite{grover1996fast}.  
During the 1990s-2010s, a series of works extended the basic idea of Grover’s algorithm to searching databases ~\cite{terhal1998single,boyer1998tight,patel2001quantum,tsai2002quantum,imre2004generalized,ju2007quantum}.  
Moreover, several attempts have been made to build full relational ``quantum DBMSs’’ in which  SQL‑like statements are realized by quantum circuits, supporting relational operations such as joins and set operations (intersection, union, difference) \cite{cockshott1997quantum, gueddana2010optimized, salman2012quantum, pang2013quantum, gueddana2014cnot, joczik2020quantum}, and data manipulation operations (insert, update, delete)  \cite{gueddana2010optimized, younes2013database, gueddana2014cnot}.

However, these approaches do not translate into practice in terms of algorithmic cost and in physical implementability. 
First, each query contains a quantum measurement, which destroys the qubit state, so the qubit state must be rebuilt for the next query, resulting in \(\mathcal{O}(N)\) time, canceling Grover’s \(\sqrt{N}\) speed‑up. %~\cite{gueddana2014cnot}. 
Second, NISQ quantum computers are noisy and limited in qubits. 
The circuit depths and number of qubits required in the above proposals, 
% demand deep strings of multi‑controlled CNOT gates on a large number of qubits
even for small relational tables, are beyond the capabilities of NISQ hardware.
% Third,   data loading is no better: preparing the initial superposition still requires \(\Theta(N)\) classical‑to‑quantum transfers, erasing any asymptotic advantage.
% \begin{enumerate}[wide,labelwidth=!,labelindent=0pt,itemsep=2pt]
%   \item \textbf{Destructive measurement.}  
%         Each query ends with a projective measurement that collapses the superposition representing the table.  
%         Re‑preparing the same state for the next query costs \(\mathcal{O}(N)\) time and resets any \(\sqrt{N}\) advantage~\cite{gueddana2010optimized,gueddana2014cnot,arxiv2405.12511}.
%   \item \textbf{Circuit depth and fan‑in.}  
%         Even modest tables require tens of qubits and sequences of \(n\)-controlled CNOTs.  
%         Current noisy‑intermediate‑scale‑quantum (NISQ) devices provide at most a few hundred physical qubits with gate infidelities of \(10^{-3}\!-\!10^{-2}\) and coherence windows of \(\mathcal{O}(10^2)\,\mu\text{s}\) (superconducting) or probabilistic entangling gates with success rates \(\le 70\%\) (photonic)~\cite{turn0search4,turn0search1,turn4search0}.  
%         The required depth therefore exceeds the error budget by orders of magnitude.
%   \item \textbf{Data‑loading overhead.}  
%         Preparing an \(n\)-qubit superposition that encodes arbitrary classical data still needs \(\Theta(N)\) classical‑to‑quantum transfers; no known QRAM scheme removes that bottleneck.
% \end{enumerate}
%
Thus, in this work,  we do \emph{not} pursue Grover‑style quantum databases.  
We prioritize quantum techniques that are technologically feasible given current constraints on qubit count and noise. As discussed in Sec.~\ref{ssec:guaranteesystems}, using the schema in Example~2, a database with 3 million rows requires only 22 qubits, which is achievable with today’s NISQ quantum processors.
% Instead, our design targets quantum advantages in \emph{privacy}, while explicitly considering the NISQ hardware's constraints and employing quantum technologies that are pocess technological feasibility for the limited qubit count and noise. as discussed in Sec.~\ref{ssec:guaranteesystems}, with the schema in Example 3, 3-million rows we just need 22 qubits which is possible with today's NISQ quantum processors. 
% in circuit depth, qubit count, and noise.\fg{the last sentence feels a bit overselling. We do not mention any of these parameters in the paper? Would be good if we did though.}

% \subsection{Private Databases}

\subsection{Classical PIR and SPIR}
\label{ssec:pqd}

Private information retrieval (PIR)
was first discussed in \cite{chor1998private, kushilevitz1997replication}. 
% PIR was first discussed in multi-server setting \cite{chor1998private}
% and single-server PIR  \cite{kushilevitz1997replication}. %was first discussed by Kushilevitz and Ostrovsky
The main challenges typically fall into two categories: designing communication-efficient approaches~\cite{lipmaa2005oblivious, cachin1999computationally, chase2021amortizing, dottling2019trapdoor} and computation-efficient approaches~\cite{melchor2016xpir, angel2018pir, ahmad2021addra}. To further reduce computation cost, recent works apply fully homomorphic encryption \cite{ali2021communication, OnionPIR, menon2022spiral} or perform partial computation in advance \cite{SimplePIR}. 
% In the multi-server setting, 
% the state-of-the-art %distributional-PIR
Recent work~\cite{DistributionalPIR} also supports complicated scenarios with skewed query distributions, where some servers receive more queries than others. 
% Crucially, the client’s choice of which server to query must remain independent of the success or failure of its previous queries. 

PIR has been studied in several concrete data management settings.  Ghinita et al.~\cite{ghinita2008private} use PIR techniques for location-based services, enabling private location queries without trusted anonymizers.  
Pantheon~\cite{ahmad2022pantheon} cuts single-round, server-side latency in key-value stores to two million item lookups in about one second.  
The tutorial~\cite{ahmad2023pirlarge} surveys both theoretical developments and practical engines for PIR over public datasets.  
$S^2$~\cite{sharma2023s2} extends multi-server, information-theoretic PIR and supports keyword search and retrieval of rows obliviously over relational tables.  
Femur~\cite{zhang2025femur} lets clients trade bounded privacy loss for better performance in public key-value stores.   
%
% Application examples of PIR include private media retrieval \cite{angel2016unobservable}, Popcorn \cite{popcorn2016nsdi} (), Peer2PIR \cite{peer2pir2025}, and HADES \cite{hades2025vldb} show system-level applications of PIR to public data services.

Symmetric private information retrieval (SPIR) provides \emph{two-sided} privacy, both data and user privacy, preventing either party from learning more than necessary  \cite{gertner1998protecting}.    
% The aforementioned systems in this section do not deliver end-to-end SPIR. 
In the classical, information-theoretic setting, SPIR is achievable only with \emph{multiple} servers that share common randomness (a shared random string) unknown to the user \cite{gertner1998protecting}.
% ; the servers do not interact during the protocol  
% Recently, practical %\emph{single-server} 
% SPIR has become feasible via homomorphic encryption~\cite{lin2022xspir}. 
% Recent work XSPIR \cite{XSPIR} presents an efficient Ring-LWE–based SPIR scheme that leverages homomorphic encryption to provide both user and database privacy with performance comparable to leading PIR systems. However, in general, classical SPIR approaches remain network- and crypto-heavy.
Recent work, XSPIR \cite{XSPIR}, presents an efficient Ring-LWE-based SPIR scheme that leverages homomorphic encryption to provide both user and database privacy, achieving performance comparable to leading PIR systems. However, classical SPIR solutions generally remain communication- and computation-heavy.

In contrast, quantum communication can, in principle, achieve SPIR with a single server. 
Our proposed quantum database tries to achieve two-sided privacy via quantum QRAC/MUB encoding and destructive measurement. 

% \emph{Quantum SPIR} By contrast, quantum protocols can eliminate the need for shared randomness in the multi-server model \cite{kerenidis2004quantum}. Moreover, quantum private queries (QPQ) \cite{giovannetti2008quantum} offer cheat-sensitive privacy and have been demonstrated experimentally using quantum key distribution (QKD) systems \cite{jakobi2011practical}. 

\subsection{Quantum SPIR} 
\label{ssec:qspir}
For completeness, we briefly outline recent advances in quantum protocols to contextualize our work within the broader quantum landscape. Readers focused solely on classical database systems may safely skip this section. 

In the \emph{multi-server} setting, quantum protocols enable sublinear communication \emph{without shared randomness}, which is unachievable in classical SPIR~\cite{kerenidis2004quantum}.
For the \emph{single-server} case,  
Quantum private query (QPQ) addresses symmetric PIR in the quantum regime. 
QPQ offers \emph{cheat-sensitive}, two-sided privacy with exponential gains in communication and computation over classical SPIR protocols~\cite{giovannetti2008quantum, GLM10TIT}. It employs two quantum states: one state for encoding the query, and a superposition decoy state to detect server-side cheating, while accessing the database via a quantum oracle. The server returns two items using only $O(\log N)$ qubits, avoiding the need to send the full $O(N)$ database, where $N$ is the database size.
A proof-of-concept experiment~\cite{de2009experimental} employed linear optics, using photon momentum and polarization as address and bus qubits.\looseness=-1

Due to challenges in scaling the dimension of the oracle operation for large databases in the original QPQ, more practical variants have been explored. A \emph{loss-tolerant} QPQ protocol based on quantum key distribution (QKD) was demonstrated in~\cite{jakobi2011practical}, with communication-efficient extensions proposed in~\cite{Gao12OE}. Other variants include high-dimensional QPQ supporting multi-bit retrieval~\cite{Wei14SciRep}, counterfactual QKD-based QPQ~\cite{Zhang13PRA}, and passive round-robin differential phase-shift QKD implementations~\cite{Li16SciRep}. An implementation over a 12.4~km fiber link incorporating error correction was demonstrated in~\cite{Chan14SciRep}.

Recent work further refines practicality and privacy. A protocol ensuring zero failure probability and bounded database information leakage is introduced in \cite{Liu15SciChina}. Decoy-state methods resilient to loss and noise were proposed in~\cite{Liu24PRA}, and extended in~\cite{Qin24PhysA} to mitigate multi-photon attacks.

From a theoretical standpoint, a relativistic QPQ protocol leveraging Minkowski causality was proposed in~\cite{sun2015relativistic}. A \emph{device-independent} QPQ approach using CHSH tests \cite{PhysRevX.3.031006} was introduced in~\cite{Maitra17PRA}, removing trust assumptions on the quantum devices.

In summary, first, most single‑server QSPIR approaches are built on quantum private query~\cite{giovannetti2008quantum}.  Instead, our private quantum database employs quantum encoding (i.e., QRACs over mutually unbiased bases), which is a new type of protocol not previously explored for the QSPIR setting. 
Second, QPQ provides cheat‑sensitive user privacy and only bounded (i.e., non‑perfect) data privacy.   In our design, data privacy is parameter‑bounded (e.g., by the copy count $k$) and thus imperfect, whereas user privacy is ensured natively by the protocol.
% \fg{I don't know any of these existing quantum approaches, but if you could strengthen difference with them and need/beaty of QRAC approach, that would be awesome}

\subsection{Complementary to Data Confidentiality and MPC}
\label{ssec:others}
Our approach is \emph{complementary} to two established privacy and security techniques: data confidentiality and secure multi-party computation (MPC). Our focus is on single-server SPIR, which ensures both user privacy (the server learns nothing about the query) and data privacy (the client learns only the requested record). In contrast, data confidentiality addresses a different problem, aiming to protect data from the server itself. Similarly, MPC enables joint computation among mutually distrustful parties. Although these techniques tackle different problems  from ours, our approach can be integrated with them to form a broader privacy-preserving data architecture.

\medskip
\para{Data Confidentiality} 
% \fg{this part seems less relevant. It is ok to keep, it just didn't tell me anything in relation to the current approach}
Existing encrypted databases focus on data confidentiality: when users move their workload to the cloud, they want to keep their data confidential to the cloud, i.e.,  the DBMS on the server executes queries without revealing unnecessary information. 
% the server (or its operator) shouldn’t learn plaintext values. They typically do not guarantee user privacy and will leak access patterns unless combined with oblivious execution.  
Hacig\"um\"us et~al.~\cite{hacigumus2002encrypted} tackle the challenge of maintaining data confidentiality on untrusted cloud providers and propose processing SQL queries over encrypted cloud databases.
MONOMI~\cite{tu2013monomi} splits analytical queries between an encrypted cloud server and an unencrypted client engine for higher efficiency. 
Performance can be further improved with secure hardware, such as IBM cryptographic coprocessors in TrustedDB \cite{bajaj2011trusteddb}. Many systems rely on %fully 
homomorphic encryption \cite{HEDA, HE3DB, ArcEDB},  
often causing high computation overhead. 
Arx \cite{Arx} takes a different path, using only semantically secure encryption schemes. 
More recent works employ differential privacy for private SQL processing~\cite{kotsogiannis2019privatesql}, or formalize leakage-safe design rules across heterogeneous cryptographic primitives~\cite{zhang2024secure}.
% These systems hide data values, yet the access pattern, namely, which tuples a user reads, often remains visible.
% Enclave ~\cite{sun2021enclave} leverages Intel SGX to build enclave-native storage
% engines for encrypted databases
%  Enclave leverages Intel SGX. Other implementation on top secure hardware include IBM cryptographic coprocessors in TrustedDB~\cite{bajaj2011trusteddb} and FPGAs in Microsoft Cipherbase~\cite{arasu2013cipherbase}.

Another promising technology for improving the security of databases outsourced to the cloud is secure enclaves, which are a hardware-based isolation mechanism recently introduced in modern CPUs such as Intel SGX \cite{mckeen2013innovative} and AMD Memory Encryption \cite{kaplan2016amd}. 
To guarantee the access-pattern privacy (i.e., hiding which tuples, blocks, or operators are touched), systems are built to support data storage \cite{sun2021enclave}, query processing \cite{Opaque, arasu2013cipherbase, ObliDB}, ACID transactions \cite{Obladi}, search index \cite{Oblix}, and fault tolerant \cite{QuORAM} in an oblivious manner. 
% Another important aspect of improving the security of databases outsourced to the cloud is

 % These systems can prevent the infrastructure from learning patterns, but still do not provide PIR-style user-privacy by default, and they remain one-sided with respect to what the client can learn.
% These systems address data confidentiality and, with oblivious plans/ORAM, access-pattern privacy, but they do not achieve PIR’s user privacy nor SPIR’s two-sided guarantee; our design complements them by pairing an oblivious feeder (classical path) with QRAC/MUB + single destructive measurement (quantum path) to provide single-server, two-sided privacy.

\medskip
\para{Secure multi-party computation}
Beyond single-server PIR, compiler-based and MPC-style frameworks achieve collaborative analytics among multiple distrustful organizations. 
SMCQL \cite{bater2017smcql} tackles the problem of processing and optimizing secure SQL queries over a federated database with mutually distrustful data owners. The queries are sent and executed by an honest broker, and thus are oblivious to others except the broker and the querier. 
To scale relational analytics queries, Conclave~\cite{Conclave} serves as a query compiler that enables a hybrid execution pipeline combining cleartext processing with secure multiparty computation, minimizing the computational overhead incurred by MPC operations. 
Secure multiparty computation is employed for collaborative analytics on secret-shared data in the cloud: SECRECY~\cite{liagouris2023secrecy} introduces oblivious relational operators and a cost model, while SCQL~\cite{fang2024secretflow} implements an industrial-grade secure query platform.  
% Although classical SPIR can be implemented using general secure two-party computation techniques, existing works typically design PIR/SPIR protocols specifically for pure retrieval tasks, as these are simpler and significantly more efficient. 
Our proposed quantum database can be integrated into existing MPC frameworks, for example, by serving as a private retrieval step. 

\section{Conclusion}
We position privacy as one of the most immediate and impactful benefits that quantum computing can offer to future database technologies. We propose a quantum database design that aims at both user and data privacy by encoding relational tuples as quantum random access codes over mutually unbiased bases. Besides this protocol, we outline a hybrid approach that embeds quantum primitives into existing DBMS engines, paving the way toward practical, application-driven quantum database systems.

\section*{AI-Generated Content Acknowledgment}
We used  OpenAI ChatGPT to polish the language and improve the readability of certain parts of this manuscript. The tool was employed for stylistic refinement, searching references, and format adjustment. All ideas, technical content, and conclusions were produced and verified by the authors.

 \bibliographystyle{plain}
% \bibliography{refCIDR.bib}
\bibliography{ref.bib}

% \clearpage
% % \input{letter}

% \clearpage
%  % \input{appendix}

% %\onecolumn
% % \appendix
% % \input{appendix.tex}

\end{document}